\documentclass[aps,prl,twocolumn,groupedaddress]{revtex4}
\usepackage{graphicx}
\begin{document}
\title{Quantum X-waves in Kerr media and the progressive undistorted squeezed vacuum}
\author{Claudio Conti}
\email{c.conti@ele.uniroma3.it}
\affiliation{%
Nonlinear Optics and Optoelectronics Labs (N{\it OO}EL)\\
National Institute for the Physics of the Matter (INFM) - Roma Tre
}%

\date{\today}

\begin{abstract}
A quantum theory of 3D X-shaped optical bullets in Kerr media is presented.
The existence of progressive undistorted squeezed vacuum is predicted.
Applications to quantum non-demolition experiments, entanglement and
interferometers for gravitational waves detection are envisaged.
\end{abstract}

\maketitle
The standard for a non-monochromatic 
quantum light packet is a superposition of photons with 
different angular frequencies and wave-vectors.\cite{LoudonBook}  This 
approach has been adopted for dispersive and 
nonlinear media, including all the spatial dimensions
(see e.g. 
\cite{Kennedy88,Deutsch91,Huttner92,Dung98,Drummond99}).
However, its application to particle-like three dimensional objects, 
like self-trapped optical bullets,\cite{TrilloBook,KivsharBook} is not trivial.

Conversely,  1D optical solitons propagating in a fiber can be
completely described at a quantum level. 
\cite{Kaup75,Carter87,Lai89,Haus90,Wright91,Yao95,Crosignani95}
This circumstance led to a series of 
extensively studied macroscopic quantum effects, 
like quantum non-demolition and squeezing.
\cite{Rosenbluh91,Friberg92,Drummond93,Schmitt98,Yu01}
 Fibers solitons are one-dimensional objects, with transversal profile given 
by the waveguide mode,
their propagation invariant nature has favored
long range
interactions, up to the recent investigation of
entangled pulses for quantum communications and computing.
 \cite{Silberhorn01}

The extension of these results to multi-dimensional
self-trapped wave-packets is an interesting enterprise.
Classically, the 3D counterpart of solitons are the mentioned
optical bullets,
which propagate without diffraction and dispersion
and are 
somehow generated in a nonlinear medium. \cite{KivsharBook, TrilloBook} 
Among the various proposal 
of 3D light bullets, due to different mechanisms and 
optically nonlinear processes in bulk media, the only experimentally demonstrated are 
the so-called ``nonlinear X-waves''. \cite{Conti03NLX,DiTrapani03}
The latter have a distinctive bi-conical shape,
which appears as an ``X'' in some section plane 
containing the propagation direction.
Notably X-waves do not
need the nonlinearity for being self-sustained, while a self-action
may favor their spontaneous formation. In essence,  
3D solitary waves destroy themselves after 
exiting the nonlinear medium, conversely an X-wave may propagate undistorted even
in vacuum (it also named ``progressive undistorted wave'',
for a review see, e.g., \cite{Recami03}). It is an ideal candidate for (quantum) information
channels in air, previously investigated by Lu and He at a classical level.\cite{Lu99}

Having in mind these properties,
it is naturally argued which is (if any) 
the quantum counterpart of nonlinear-X-waves.
Is it possible to provide a fully quantized version at any photon number, 
as in the case of optical solitons? This is the subject of this Letter. 
I will report a quantum theory of 
nonlinear X-waves and some of its corollaries;
 in particular the possibility of 
generating a very exotic (but at the same time intriguing for many applications) 
state of light: the progressive undistorted squeezed vacuum.
Quantum nonlinear X-waves turn out to 
have the same properties of quantum fiber solitons:
they can propagate with a well defined photon number, thus
enabling sequential measurements on the same quantum state;\cite{Drummond93} 
all the applications of quantum fiber solitons may be translated in the 
quantum X-waves world,
with the non trivial benefits of a 3D space.

The basic idea underlying this work is using 3D wave packets 
parametrized by their velocity, instead of angular frequency, 
for a quantization procedure. 
Since it is interesting to be as much as possible near to realm of experiments,
the starting point is the equation for the motion of 
an optical pulse, with complex amplitude $A$, traveling with diffraction in a normally dispersive
medium. It can be equivalently cast as an evolution problem with respect to 
the direction of propagation $z$, or to time $t$,\cite{Kennedy88} 
which I adopt here to be consistent with standard 
quantum mechanics:
\begin{equation}
\label{main1}
i\frac{\partial A}{\partial t}+i \omega' \frac{\partial A}{\partial z}
-\frac{\omega''}{2}\frac{\partial A}{\partial z^2}+\frac{1}{2k}\nabla_{xy}^2 A
=\frac{\delta \mathcal{H}_I}{\delta A^*}\text{.}
\end{equation}
$\omega'$ and $\omega''$ are 
the first order and the modulus of the second order dispersion terms ($\omega''>0$);
$k=n(\omega_0)\omega_0/c$ and 
$\omega_0$ are the wave-number and the carrier angular frequency, respectively,
and $n=n(\omega_0)$ is the refractive index.
$\mathcal{H}_I$ is a classical interaction Hamiltonian taking into account nonlinear
effects.

In the linear case ($\mathcal{H}_I=0$) the general radially symmetric 
($r^2\equiv x^2+y^2$, in the following) solution 
can be expressed as a superposition, parametrized by the velocity $v$, of a special class of 
undistorted progressive waves, the radially symmetric X-waves, given by
three dimensional complex profile:\cite{Conti04}
\begin{equation}
\displaystyle\psi_q^{(p)}(z,r)=\displaystyle\int_0^\infty  f_q(\alpha)J_0(\sqrt{\frac{\omega'' k_0}{\omega'}}\alpha r)
e^{i(\alpha-\frac{p}{\hbar})z}d\alpha \text{,}
\end{equation}
with $f_q=\sqrt{k/\pi^2\omega'(p+1)}(\alpha\Delta)L_p^{(1)}(2\alpha\Delta)\exp(-\alpha\Delta)$ 
(related details are also given in \cite{Conti2003quantum}), $\Delta$ is parameter measuring
the transversal spatial extent of the beam,
$L_q^{(1)}$ the generalized Laguerre polynomials ($q=0,1,...$), 
and having introduced, for later convenience,
 the {\it momentum} $p=m v$ of the X-wave, ($-\infty<p<\infty$)
 with $m=\hbar/\omega''$ ($p/\hbar=v/\omega''$).
\begin{figure}
\includegraphics[width=8.3cm]{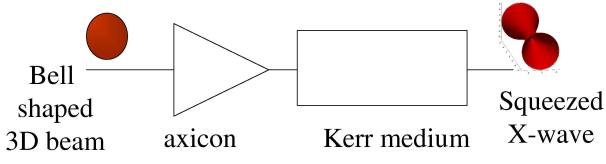}
\caption{(Color online) The squeezed nonlinear X-wave device. \label{figureDEVICE}}
\end{figure}

The resulting analytic signal for the electric field $E$ is 
\begin{equation}
\label{electricfield}
\begin{array}{l}
E=\sqrt{\frac{2}{\epsilon_0 n^2 m}}\displaystyle\sum_q \int  
\,C_q(p)\times\\
\exp[-i\omega_q(p)t+i k_0 z]\psi_q^{(p)}[z-(\omega'+\frac{p}{m})t,r]dp
\end{array}
\end{equation}
with $\omega_q(p)=\omega_0+p^2/2\hbar m$.
The electromagnetic classical energy is 
\begin{equation}
\mathcal{E}=\displaystyle\int\int\int |A|^2 dxdydz=\displaystyle\sum_q\int |C_q|^2 dp \text{.}
\end{equation}
Using this formulation a non-monochromatic pulsed beam is written as an 
integral sum of wave-packets with different velocities,
or momentum $p$, and the diffraction/dispersion
process is reduced to a one-dimensional evolution.

Eq. (\ref{electricfield}) is a superposition of harmonic oscillators,
with angular frequencies $\omega_q(p)$, each weighting a traveling mode
(i.e. corresponding to the usual cavity mode, with the difference
here that it is rigidly moving). It can be standardly quantized,
  and the resulting positive frequency field operator is 
(in the Heisenberg picture
for the Dirac operators $a_p$)
\begin{equation}
\label{electricfield1}
E=\displaystyle\sum_p \int  \sqrt{\frac{2\hbar \omega_q(p)}{\epsilon_0 n^2 m}}
 a_q(p,t)e^{i k_0 z}\psi_q^{(p)}[z-(\omega'+\frac{p}{m})t,r]dp
\end{equation}
where the energy of each elementary excitation is given by
$\hbar\omega_q(p)=\hbar\omega_0+p^2/2m$, 
and the Hamiltonian $H_0=\sum_q \int  \hbar \omega_q(p)a_q^\dagger (p)a_q(p) dp$
(the zero-point energy has been renormalized, as usual in quantum field theory \cite{ItzyksonBook}).

This suggestive way of quantizing the propagating optical field, is equivalent 
to the standard plane wave expansion. It is in essence a change of basis, which
enables to represent the 3D evolution as a 1D quantum gas of particles
with momentum $p$, and mass $m$.

In the presence of a nonlinear coupling between the quasi-particles, due to the Kerr effect,
the contribution to the 
classical energy is 
\begin{equation}
\mathcal{H}_{I}=\displaystyle \frac{\chi}{2}\int\int\int |A|^4 dx dy dz\text{,}
\end{equation}
with $\chi<0$ in a focusing medium.
The corresponding interaction Hamiltonian is, after some manipulations
and with obvious notation,
\begin{equation}
\label{HI}
\begin{array}{l}
H_I=\displaystyle\frac{1}{2}\sum_{lmno} \int \,
\sqrt{\omega_o(p_4)\omega_n(p_3)\omega_m(p_2)\omega_l(p_1)} \times
\\
\chi_{lmno}(p_4+p_3-p_2-p_1)
a_l^\dagger(p_4) a_m^\dagger(p_3) a_n(p_2) a_o(p_1) d^4 {\bf p}\text{.}
\end{array}
\end{equation} 
The interaction kernel, the ``vertex,'' $\chi_{lmno}(\nu)$ 
turns out to be the Fourier transform of the spatial transversal superposition of the 
component X-waves profiles:
\begin{equation}
\kappa_{lmno}(z)=\frac{\chi \hbar^2}{m^2}
\int \int (\psi_l^{(0)})^* (\psi_m^{(0)})^* \psi_n^{(0)}\psi_o^{(0)} dx dy\text{.}
\end{equation}
As an application, consider the ``device'' shown in figure \ref{figureDEVICE}:
an X-wave, generated by the axicon, travels in a Kerr medium.
The axicon must be intended in a generalized sense, i.e. either a 
linear device, as those typically employed in linear experiments,\cite{Saari97,Grunwald03} or a 
nonlinear process that furnishes the required spatio-temporal reshaping 
of the laser pulsed beam into an X-wave.\cite{Conti03NLX,DiTrapani03}
By this approach, only a $p-$superposition of a single basis element, denoted by index $q$,
 must be taken into account and (\ref{HI}) becomes 
[with $a_m(v)=\delta_{qm} a(v)$ and omitting hereafter the index $q$
(e.g. $\chi_{lmno}\rightarrow \chi)$]
\begin{equation}
\label{HIsimpler}
\begin{array}{l}
H_I=\displaystyle\frac{1}{2} \displaystyle\int   \sqrt{\omega(p_4)\omega(p_3)\omega(p_2)\omega(p_1)}\times
\\\chi(p_4+p_3-p_2-p_1)
a^\dagger(p_4) a^\dagger(p_3) a(p_2) a(p_1) d^4 {\bf p}\text{.}
\end{array}
\end{equation} 

The analysis is further simplified in the low-momentum approximation, typically
adopted in the physics of weakly interacting bosons.\cite{Landau} This corresponds to 
assume that the velocities are all in proximity of the linear group velocity, such that
($p_j\cong0$)$ \sqrt{\omega(p_4)\omega(p_3)\omega(p_2)\omega(p_1)}\cong \omega_0^2$.
Introducing the particle operator 
$\phi(z)=(1/2\pi\hbar)\int a(p) \exp(i p z/\hbar)dp$, the interaction 
Hamiltonian is given by 
\begin{equation}
\label{HI3}
H_I= \frac{1}{2}\int  \sigma(z) \phi^\dagger(z)\phi^\dagger(z) \phi(z) \phi(z) dz\text{,} 
\end{equation}
with
\begin{equation}
\sigma(z)=(2\pi\hbar)^3\frac{\chi \omega_0^2 \hbar^2}{m^2} \int \int |\psi^{(0)}|^4 dx dy\text{.}
\end{equation}
By expressing $a$ in terms of $\phi$ in (\ref{electricfield1}),the previous results can be reformulated as follows.
The whole 3D evolution of the electric field is given, in the Heisenberg picture for the
particle operators $\phi$ and $\phi^{\dagger}$, \cite{note1} by
\begin{equation}
\label{finalE}
E=\sqrt{\frac{2\hbar\omega_0}{\epsilon_0 n^2 m}}e^{ik_0 z-i\omega_0 t}\int  \xi(s,z,t,r)\phi(s,t) ds
\end{equation}
with 
\begin{equation}
\xi(s,z,t,r)=\displaystyle\int  \sqrt{\frac{\omega(p)}{\omega_0}} \psi^{(p)}[z-(\omega'+
\frac{p}{m})t,r]e^{-i p s} dp\text{.}
\end{equation}
The Heisenberg evolution equation for $\phi$ is the generalized nonlinear quantum Schr\"odinger equation:
\begin{equation}
\label{gQNLSE}
i\hbar \frac{\partial \phi}{\partial t}(t,z)
=-\frac{\hbar^2}{2m}\frac{\partial^2 \phi}{\partial z^2}(t,z) +\sigma(z) \phi^\dagger(t,z) \phi(t,z)\phi(t,z)\text{.}
\end{equation}

Eq. (\ref{finalE}) shows that the whole evolution is the composition of the deterministic propagation of the $X-wave$ $\psi$,
embedded in the $\xi$ kernel, and of the quantum one obeying (\ref{gQNLSE}).
Taking for $\psi$ the fundamental X-wave ($q=0$), 
$\sigma(z)$ is a bell shaped function. It is possible to show that, 
if the classical dispersion length is much smaller that the diffraction length,
it can be treated as a constant $\sigma$ and the model reduces to the integrable 
quantum nonlinear Schr\"odinger equation. \cite{Kaup75}
Hence the whole 3D quantum dynamics is reduced to an exactly solvable model. 
Conversely when the diffraction length is smaller
 than the dispersion length, the dispersion is negligible, and the model
is still integrable, representing self-phase modulation.

Summarizing, the 3D nonlinear quantum propagation of X-waves
can be treated in terms of a well known approach. 
All the experiments concerning quantum solitons, 
involving quantum non-demolition, squeezing and entanglement can be 
re-stated in terms of undistorted progressive 3D wave-packets.
Quantum non-demolition experiments by collision of X-waves with different velocities,
generated by different axicons, can be envisaged and analyzed with 
the same techniques previously developed, which will not be reported here
(for a review see \cite{Drummond93} and references therein).

\begin{figure}
\includegraphics[width=8cm]{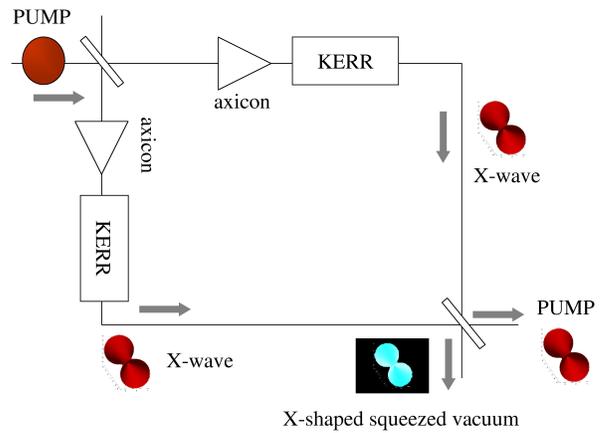}
\caption{(Color online) X-wave nonlinear Mach-Zehnder interferometer for the generation of the squeezed vacuum. \label{figureMZ}}
\end{figure}

In figure \ref{figureMZ},
an idealized interferometer for the generation of squeezed 
nonlinear X-waves is sketched. More elaborated setups may be readily
drawn from previously developed fiber solitons schemes.\cite{Yu01}
From a single bell shaped 3D wave-packet in air, two 
identical X-waves are generated, propagate in a Kerr medium, and 
then interfere at a symmetric beam splitter.
At one output port a squeezed X-wave is obtained; the degree of squeezing 
can be monitored by balanced photo-detection. \cite{Shirasaki90}
With the modifications used for fiber solitons,\cite{Silberhorn01} 
entangled 3D (progressive undistorted) pulsed beams can be generated. 

The electromagnetic field attained at
the other output port is peculiar. It is the so-called squeezed vacuum, \cite{Bergman91} 
which can be interpreted as the nonlinearly induced quantum noise, 
``dissected'' from the pump beam,\cite{Drummond93} with no average electric field but with 
spatio-temporal correlation. 
Furthermore it has 
the additional, remarkable, property of having the propagation
characteristics of a progressive invariant 3D wave. 
Hence its spectral properties are X-shaped, and can be distinguished 
inside standard vacuum fluctuations. 
This property makes such a state appealing for applications in interferometry, 
where the squeezed vacuum can be used for enhancing the performances, 
\cite{Caves81} 
as in the framework of the LIGO (or even MIGO \cite{Chiao2003}) projects for the
detection of gravitational waves, \cite{Kimble01,Buonanno2004} 
or for geophysical studies. \cite{Grishchuk04}
 To achieve sensitivity below the so-called
quantum limits, it has been proposed to use squeezed vacuum  
as input in one of the arms of the interferometer used to detect gravitational waves. 
The effective absence of 
diffraction for progressive undistorted waves may provide the elements
for a reduction of the very high power levels needed to reduce shot noise. 
Furthermore the X-shaped squeezed vacuum is expected to be very robust with 
respect to the contamination of the standard vacuum fluctuation, a significant problem
 in the future generations of LIGO. In some sense, the spatio-temporal modulation, 
characterizing the progressive invariant waves, may be used to encode useful signals and 
discern them from noise.

In conclusion, the quantum propagation of nonlinear X-waves has been investigated.
All the quantum effects, which have been previously considered for fiber quantum solitons, 
do have a counterpart in the 3D realm of progressive undistorted packets.
Thus, a variety of new experiments may be conceived, with applications  
ranging from quantum information to gravitational waves detection.

I thank G. Assanto, A. Ciattoni, B. Crosignani, E. Recami and S. Trillo for the inestimable discussions.

\begin{thebibliography}{39}
\bibitem[{not()}]{note1}
\bibinfo{note}{The carrier frequency $\exp(-i\omega_0 t)$ has been factorized
  out from $\phi$}.
\end{thebibliography}

\end{document}